\documentclass[a4paper]{jpconf}
\usepackage{graphicx}
\begin{document}
\title{Searches for the Violation of Pauli Exclusion Principle at LNGS in VIP(-2) experiment}

\author{H.~Shi$^1$, 
S.~Bartalucci$^1$, 
S.~Bertolucci$^2$, 
C.~Berucci$^{3,1}$, 
A.M.~Bragadireanu$^{1,4}$, 
M.~Cargnelli$^3$, 
A.~Clozza$^1$, 
C.~Curceanu$^{1,4,5}$, 
L.~De Paolis$^1$, 
S.~Di Matteo$^6$,
A.~d'Uffizi$^1$,
J.-P.~Egger$^7$,
C.~Guaraldo$^1$,
M.~Iliescu$^1$,
T.~Ishiwatari$^2$,
J.~Marton$^2$,
M.~Laubenstein$^8$,
E.~Milotti$^9$,
D.~Pietreanu$^{1,4}$,
K.~Piscicchia$^{1,5}$,
T.~Ponta$^4$,
A.~Romero Vidal$^{10}$,
E.~Sbardella$^1$,
A.~Scordo$^1$,
D.L.~Sirghi$^{1,4}$,
F.~Sirghi$^{1,4}$,
L.~Sperandio$^1$,
O.~Vazquez Doce$^{11}$,
E.~Widmann$^2$ and 
J.~Zmeskal$^2$
}

\address{$^1$INFN, Laboratori Nazionali di Frascati, Via E. Fermi 40, I-00044 Frascati(Roma), Italy}
\address{$^2$CERN, CH-1211, Geneva 23, Switzerland}
\address{$^3$Stefan-Meyer-Institut f\"{u}r Subatomare Physik, Boltzmanngasse 3, 1090 Wien, Austria}
\address{$^4$IFIN-HH, Institutul National pentru Fizica si Inginerie Nucleara Horia Hulubbei, Reactorului 30, Magurele, Romania}
\address{$^5$Museo Storico della Fisica e Centro Studi e Ricerche Enrico Fermi, Piazza del Viminale 1, 00183 Roma, Italy}
\address{$^6$Institut de Physique UMR CNRS-UR1 6251, Universit\'{e} de Rennes1, F-35042 Rennes, France}
\address{$^7$Institut de Physique, Universit\'{e} de Neuch\^{a}tel, 1 rue A.-L. Breguet, CH-2000 Neuch\^{a}tel, Switzerland}
\address{$^8$INFN, Laboratori Nazionali del Gran Sasso, S.S. 17/bis, I-67010 Assergi (AQ), Italy}
\address{$^9$Dipartimento di Fisica, Universit\'{a} di Trieste and INFN-Sezione di Trieste, Via Valerio, 2, I-34127 Trieste, Italy}
\address{$^{10}$Universidade de Santiago de Compostela, Casas Reais 8, 15782 Santiago de Compostela, Spain}
\address{$^{11}$Excellence Cluster Universe, Technische Universit\"{a}t M\"{u}nchen, Boltzmannstra\ss e 2, D-85748 Garching, Germany}

\ead{hexishi@lnf.infn.it}

\begin{abstract}
The VIP (Violation of Pauli exclusion principle) experiment and its follow-up experiment VIP-2 at the Laboratori Nazionali del Gran Sasso (LNGS) 
search for X-rays from Cu atomic states that are prohibited by the Pauli Exclusion Principle (PEP). 
The candidate events, if they exist, will originate from the transition of a $2p$ orbit electron to the ground state which is already occupied by two electrons. 
The present limit on the probability for PEP violation for electron is 4.7 $\times10^{-29}$ set by the VIP experiment. 
With upgraded detectors for high precision X-ray spectroscopy, 
the VIP-2 experiment will improve the sensitivity by 2 orders of magnitude. 
\end{abstract}

\section{Introduction}
In quantum mechanics the Pauli exclusion principle can be formalized with two fundamental principles. 
First, the states of identical particles are described in terms of wave functions. 
Second, with respect to the permutation of identical particles, the states are necessarily either all symmetrical for bosons, 
or all antisymmetrical for fermions. 
The second part contains the ``symmetrization postulate'' \cite{Mes62}, which excludes mixing of different symmetrization groups. 
Messiah and Greenberg noted in \cite{Mes64} that this superselection rule ``does not appear as a necessary feature of the QM description of nature.''
In this context, the violation of PEP is equivalent to the violation of statistics \cite{Gre00},
and experimentally to the existence of states of particles that follow 
the statistics other than the fermion or the boson ones.

Exhaustive reviews of the experimental and theoretical searches for a small violation of the PEP or the violation of statistics can be found for example in \cite{Ell12} and \cite{Gre00}.
We point out that firstly, there is no established model in quantum field theory that can include small violations of the PEP explicitly.
Secondly, although many experimental searches present limits for the violation, 
the parameters that quantify the limits are model/system dependent and are not generally comparable. 
Moreover, to search for states that are in an abnormal or a mixed symmetry, 
it is crucial to introduce into the system an abundance of new states, 
among which the abnormal states may be found. 
Ramberg and Snow \cite{Ram90} took this argument into account, 
by running a high electric current through a Cu conductor, 
and they searched for X-rays from transitions that are PEP-forbidden after one electron is captured by a Cu atom. 
Due to the shielding effect of the additional electron in the ground level, 
the energy of such abnormal transitions will deviate from the Cu K$\alpha$ X-ray at 8 keV by about 200 eV \cite{Cur13}, 
which will be distinguishable in precision spectroscopy. 
Since the $new$ electrons from the current are supposed to have no established symmetry with the electrons inside the Cu atoms, 
the detection of the abnormal X-ray is an explicit indication of the violation of statistics, and thus the violation of the PEP.

We want to mention that, 
one known system in which the dichotomy of fermions and bosons does not work is in the 2-dimensional condensed matter physics through the (fractional) Quantum Hall effect \cite{Pra90}.
Particles that are neither fermions nor bosons, and that may exist in electronic systems confined to two spatial dimensions have been constructed theoretically and 
investigated in the laboratory with great consistency with the theories as reviewed in \cite{Ste08}. 
The physics of this special system is exciting in itself and may provide hints to the searches for the violation of the exclusion principle in other systems.

\begin{table}[htbp]
\caption{
         The improvement factors for VIP2 in comparison to the features of VIP \cite{Mar13} : 
        }
\label{tab:improvement}
\begin{center}
\begin{tabular}[b]{l l l}
\hline
\hline
Changes in VIP2     &  value VIP2 (VIP)  &    expected gain  \\
\hline
acceptance          & 12 \% ($\sim$ 1 \%)     &  12  \\
increase current    & 100 A (40 A)    & $>$ 2   \\
reduced length      & 3 cm (8.8 cm )  &  1/3 \\
\hline
total linear factor &              &     8  \\
\hline 
energy resolution &  170 eV (320 eV) @ 8 keV   &  4 \\   
reduced active area  & 6 cm $^2$ (114 cm $^2$)   &   20 \\
better shielding and veto   &         &    5-10 \\
higher SDD efficiency   &   &    1/2 \\
\hline 
background reduction  &  &  200 - 400 \\
\hline 
overall improvement  &  &   $>$ 120 \\
\hline
\end{tabular}
\end{center}
\end{table}

\section{VIP-2 Experiment}

The VIP experiment used a similar method to the Ramberg-Snow experiment, 
and the same definition of the parameter to represent the probability that the PEP is violated for a direct comparison of the experimental results. 
An improvement in sensitivity was achieved 
firstly by performing the experiment in the low radioactivity laboratory at LNGS, 
which has the advantage of the excellent shielding against cosmic rays. 
Secondly the application of Charge Coupled Device (CCD) as the X-ray detector with a typical energy resolution of 320 eV at 8 keV, 
increased the precision in the definition of the region of interest to search for anomalous X-rays. 
The VIP experiment set a limit for the probability of the PEP violation to be 4.7 $\times 10^{-29}$ \cite{Bar06, Bar09}.

By introducing new X-ray detectors and an active shielding, as listed in the Table \ref{tab:improvement}, 
the VIP-2 experiment expects to further improve the sensitivity by two orders of magnitude. 
The detectors and the setup of the VIP-2 experiment are introduced by more detail in 
\cite{Pic16} from this conference proceedings and in \cite{Shi15, Mar13},

\section{Status of preparation}
From 2014 to 2015, we finished the production of 32 pieces of veto detectors made of plastic scintillators coupled to Silicon PhotoMmultipliers(SiPM). 
The preamplifiers for the SiPMs developed by the Stefan-Meyer-Institut (SMI) in Vienna, 
were fine tuned and with the veto detectors we took background data inside the laboratory. 
The six SDD elements with a total active area of 6 cm$^2$ were mounted inside the vacuum chamber and cooled down to 110 K, 
and we successfully took calibration data using an Fe-55 source. 
All the SDDs were performing according to expectation with an energy resolution of about 150 eV (FWHM) at 6 keV. 

\begin{figure}[htbp]
\centering
\includegraphics[width=14cm,clip]{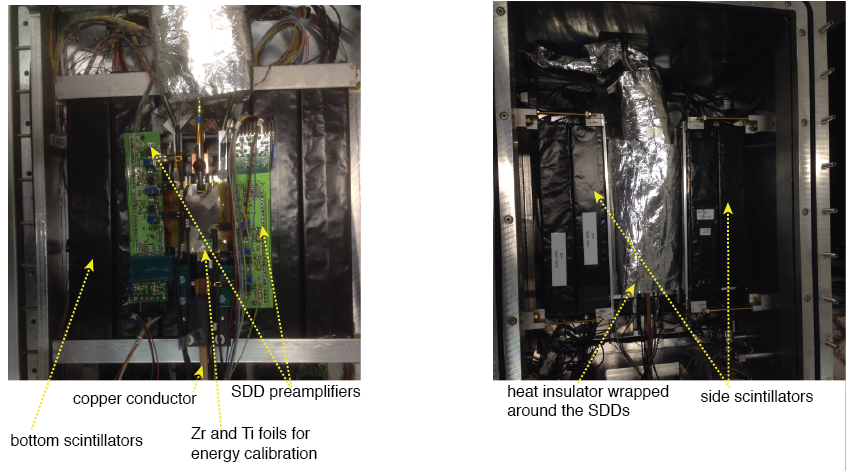}
\caption{Pictures of the detectors inside the setup. 
         Left : a top view of the setup after the SDDs and their preamplifier board were mounted close to the Cu conductor;
         the bottom layer of scintillators is also visible in the background; 
         right : same angle of view after the heat insulator was wrapped around the SDDs, and the side layer of scintillators were mounted. 
         }
\label{fig:detectors}       
\end{figure}

We define the trigger for data taking by either an event at any SDD or a coincidence between two layers of the veto detector. 
The trigger logic was implemented using the NIM standard modules and a VME-based data acquisition system was customized.
From the cosmic ray events that produce coincidences between the veto scintillators and the SDDs,
we confirmed the time correlation whose spread is characterized by the drift time of the SDD of less than 1 $\mu s$. 
The SDD spectra and the timing correlation plots are shown in \cite{Pic16} of this conference proceedings. 

To make sure the heat from the Cu conductor when the electric current is applied does not affect the silicon detectors nearby,
we monitored the temperatures of the setup while applying an electric current up to 80 Ampere. 
A water chiller with cooling capacity less than 900 W was confirmed to be sufficient to keep the Cu at the room temperature level. 

\begin{figure}[htbp]
\centering
\includegraphics[width=14cm,clip]{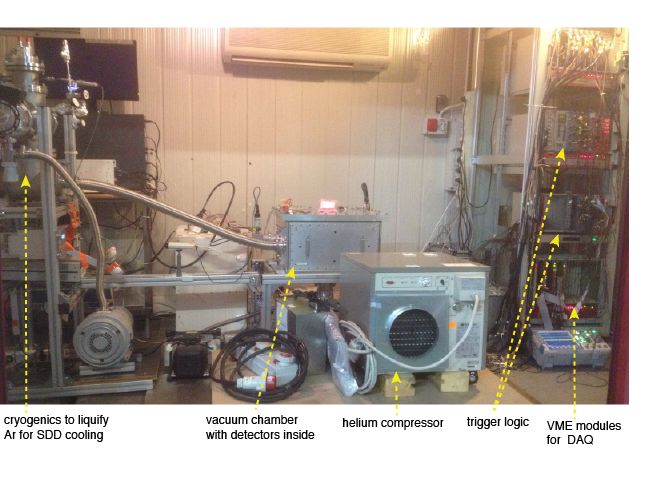}
\caption{
         A picture of the VIP-2 setup in the barrack at LNGS as in November 2015. 
         }
\label{fig:barrack}       
\end{figure}

The series of tests for the detectors, the trigger logic and the slow control/monitors 
have confirmed that the expected performances of the apparatus, as listed in Table \ref{tab:improvement}, were reached. 
In November 2015, we transported the full setup from Vienna to the VIP-2 experiment site in Gran Sasso. 
We assembled the cryogenic system and the readout logic as shown in Fig. \ref{fig:barrack} and started the test run.

\begin{figure}[htbp]
\centering
\includegraphics[width=15cm,clip]{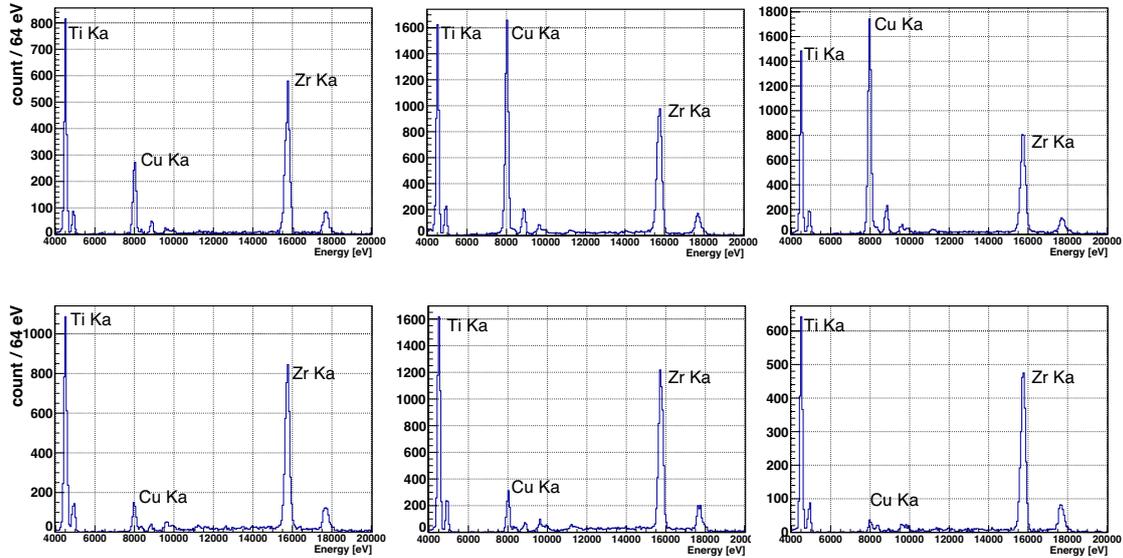}
\caption{
        Energy spectra of SDDs from a calibration run in the VIP barrack at LNGS. 
        The relative intensity of the Cu Ka varies due to geometrical reason. 
        }
\label{fig:spectra}       
\end{figure}

As an important step to confirm the operation of the SDDs after the transportation, 
we took calibration data 
by shining 22 keV X-rays on Zr and Ti foils placed near the SDDs. 
The spectra obtained are shown in Fig. \ref{fig:spectra}, 
where all SDDs were in working condition and a preliminary analysis showed that the energy resolution is compatible to the one obtained in the tests in the SMI laboratory. 

\section{Summary and perspectives}
We are searching for possible violation of the Pauli exclusion principle for electrons by improving the method originally proposed by Ramberg and Snow.
The application of high resolution X-ray detectors with timing capability is the key factor to improve the sensitivity. 

In year 2015, we finished the productions of all the detectors and the apparatus of the VIP-2 experiment, 
and transported the whole setup to LNGS. 
From the test run in December 2015, 
the detectors showed same performance as confirmed in previous tests.

We started a data taking of about 3 years with the aim to either improve by two orders of magnitude 
the probability of the PEP violation by electrons, or to find a small violation.

\section*{Acknowledgements}
We thank H. Schneider, L. Stohwasser, and D. St\"{u}ckler from Stefan-Meyer-Institut 
for their fundamental contribution in designing and building the VIP2 setup. 
We thank the very important assistance of the INFN-LNGS laboratory staff during all phases of preparation, 
installation and data taking as well as the support from the HadronPhysics FP6(506078), 
HadronPhysics2 FP7 (227431), HadronPhysics3 (283286) projects and the EU COST 1006 Action is gratefully acknowledged. 
Especially we thank the Austrian Science Foundation (FWF) which supports the VIP2 project with the grant P25529-N20, 
and we thank the support from the EU COST Action MP1006, Fundamental Problems in Quantum Physics, 
and from Centro Fermi (``Problemi aperti nella meccania quantistica''project).
Furthermore, this paper was made possible through the support of a grant from the John Templeton Foundation (ID 581589).
The opinions expressed in this publication are those of the authors and do not necessarily reflect the views of the John Templeton Foundation.

\end{document}